\documentclass[twoside, 11pt]{article}

\usepackage{mathrsfs,amsfonts,amsmath}
\RequirePackage{ifthen,calc}
\usepackage{theorem}
\usepackage[dvips]{color}
\usepackage{graphicx}
\usepackage{subfigure}

\def\E{{\mathbb E}}

 \catcode`@=11
 \def\@evenhead{\hbox to\textwidth{\footnotesize\rm\thepage \hfill
  {\it }}} 

 \def\@oddhead{\hbox to \textwidth{\footnotesize{\it
 } \hfill\thepage}}

 \renewcommand{\section}{\makeatletter
 \renewcommand{\@seccntformat}[1]{{\csname the##1\endcsname.}\hspace{0.45em}}
 \makeatother \@startsection
{section}
{1}
{0pt}
{\baselineskip}
{0.5\baselineskip}
{\normalsize\bfseries\mathversion{bold}}}

\renewcommand{\subsection}{\makeatletter
 \renewcommand{\@seccntformat}[1]{{\csname the##1\endcsname.}\hspace{0.45em}}
 \makeatother \@startsection
{subsection}
{1}
{0pt}
{\baselineskip}
{0.5\baselineskip}
{\normalsize\bfseries\mathversion{bold}}}

\catcode`@=12

\newtheorem{thm}{\noindent Theorem}[section]

\theoremheaderfont{\normalfont\bfseries}
\theorembodyfont{\slshape} {\theorembodyfont{\rmfamily}} {\theorembodyfont{\rmfamily}
\newtheorem{defn}{\noindent Definition}[section]}
{\theorembodyfont{\rmfamily} } {\theorembodyfont{\rmfamily}
}
{\theorembodyfont{\rmfamily} } {\theorembodyfont{\rmfamily}
}
{\theorembodyfont{\rmfamily} } {\theorembodyfont{\rmfamily} } {\theorembodyfont{\rmfamily}
}

 \def\beqlb{\begin{eqnarray}}\def\eeqlb{\end{eqnarray}}
 \def\beqnn{\begin{eqnarray*}}\def\eeqnn{\end{eqnarray*}}
 \numberwithin{equation}{section}
\def\2R{\mathbb{R}_+\times\mathbb{R}}
\def\qed{\hfill$\square$\smallskip}

\def\3R{\mathbb{R}_+\times\mathbb{R}_-}

\usepackage{indentfirst}

\begin{document}

\title{\bf Elastic Net Procedure for Partially Linear Models
\footnotetext{\hspace{-5ex}
${[1]}$ School of Mathematics, Guangxi University, Nanning 530004, China.
\\${[2]}$ School of Statistics, Shandong University of Finance and Economics.
Jinan 250014,  China.
\\${[3]}$ School of Mathematics and Statistics, Hechi University, Yizhou  546300, China
\\${[4]}$ E-mail:daihongshuai@gmail.com.
\newline
}}
\author{\small Chunhong Li$^1$,  Dengxiang Huang$^{1}$,  Hongshuai Dai$^{2,4}$, Xinxing Wei$^{1,3}$ }

 \maketitle

\begin{abstract}
Variable selection plays an important role in the high-dimensional data analysis. However the high-dimensional data often induces the strongly correlated variables problem. In this paper, we propose Elastic Net procedure for partially linear models and prove the group effect of its estimate. By a simulation study, we show that the strongly correlated variables problem can be better handled by the Elastic Net procedure than  Lasso, ALasso and  Ridge. Based on an empirical analysis, we can get that the Elastic Net procedure is particularly useful when the number of predictors $p$ is much bigger than the sample size $n$.
\end{abstract}


\bigskip
\section{Introduction}
The high-dimensional data is widely used in medical research, bioinformatics, econometrics etc. It has attracted a lot of interest recently.  Variable selection is fundamentally important for knowledge discovery with the high-dimensional data and it could greatly enhance the prediction performance of the fitted model. Traditional model selection procedures follow best-subset selection and its step-wise variants. However, best-subset selection is computationally prohibitive when the number of predictors is large, and it is unstable. Thus, the resulting model has poor prediction accuracy. To overcome these drawbacks of subset selection, statisticians have recently proposed various penalization methods to perform simultaneous model selection and estimation. In particular, the Lasso (Tibshirani\cite{T1996}) and  SCAD (Fan and Li \cite{FL2001}) are two very popular methods due to their good computational and statistical properties. Efron et al. \cite{EHJT2004} proposed the LARS algorithm for computing the entire Lasso solution path. Knight and Fu \cite{KF2000} studied the asymptotic properties of the Lasso. Fan and Li \cite{FL2001} showed that the SCAD enjoys the oracle property. But the oracle property does not hold for Lasso. Then, Zou \cite{Z2006} proposed the Adaptive Lasso (ALasso) by utilizing the adaptively weighted $l_{1}$ penalty, which has the oracle property.

Correlated variables are very important in applications and theory. So it is interesting and important to estimate coefficients of the correlated variables. However, the methods mentioned above can not deal with the strongly correlated variables perfectly. Zou and Hastie \cite{ZH2005} introduced the Elastic Net procedure which  can deal with the strongly correlated variables effectively. The essential strongly correlated variables tend to be into the model together for the group effect of the Elastic Net. Furthermore, similar to Lasso and Ridge estimation, the Elastic Net procedure has some excellent properties. Thus, it has already received a considerable amount of attention. Zou and Zhang \cite{ZZ2009} proved the oracle property of the Adaptive Elastic Net. Chen et al. \cite{CYZL2012} showed that the profiled Adaptive Elastic Net for partially linear models also has the oracle property. Its estimation identifies the right subset model and has the optimal estimation rate. But little  work has been done on the highly correlated variables. So we will investigate whether the Elastic Net encourages the group effect in partially linear models in this paper.
The paper is organized as follows. In Section 2, we turn  partially linear models into classical linear models by the kernel estimation. The Elastic Net procedure for partially linear models is presented in this section as well. In Section 3, we discuss the group effect that is caused by the Elastic Net penalty for partially linear models. The simulation results comparing  Lasso, ALasso,  Ridge and the Elastic Net are presented in Section 4. Section 5 studies a real date example.

\bigskip

\section{Elastic Net procedure}
Partially linear models  are a class of commonly-used semiparametric models, which are flexible enough and well interpretable, since they contains both parametric and nonparametric components.
Next, we consider the Elastic Net procedure for partially linear models and make a further study of its group effect.

Consider the following partially linear model,
\begin{equation}\label{formula3}
 \ Y=X'\beta+f(T)+\varepsilon,
\end{equation}
where  $X=(x_1,\cdots,x_p)'$, $\beta=(\beta_1,\cdots,\beta_p)'$ is sparse which means that only some components are nonzero, and $f(\cdot)$ is an unknown smooth function of the covariate $T$, $\varepsilon$ is random error  with expectation 0 and the standard deviation $\sigma$, which is independent of $(X,T)$. In this paper, we only consider univariate $T$.
From \eqref{formula3}, we have
$$f(T)=\E(Y|T)-\E(X|T)'\beta.$$
Then
\begin{equation}\label{formula4}
\ Y-\E(Y|T)=\big(X-\E(X|T)\big)'\beta+\varepsilon.
\end{equation}

Obviously, we can turn the partially linear model into the classical linear model if $\E(X|T)$ and $\E(Y|T)$  are known. We  estimate $m_X(T)=\E(X|T)$ and $m_Y(T)=\E(Y|T)$ by the kernel estimation.
 Suppose a random sample of $n$ individuals is chosen.
Let ${\bf X}=(X'_1,X'_2,...,X'_n)'$ be the design matrix, where $X_i=(x_{i1},x_{i2},...,x_{ip})'$, $i=1,2,...,n$. Similarly, we assume that ${\bf Y}=(y_1,y_2,...,y_n)'$, ${\bf T}=(T_1,T_2,...,T_n)'$,
 and ${\bf \epsilon}=(\varepsilon_1,\varepsilon_2,...,\varepsilon_n)'$. Moreover, denote the estimators of $m_X(T)$ and $m_Y(T)$ by $\hat{m}_X(T)$ and $\hat{m}_Y(T)$, respectively. Then,
  $$\hat{m}_X(T)=\frac{\displaystyle\sum_{i=1}^n{K(\frac{T_i-T}{h})}}{\displaystyle\sum_{i=1}^n{K(\frac{T_i-T}{h})}}X_i,$$
  and  $$\hat{m}_Y(T)=\frac{\displaystyle\sum_{i=1}^n{K(\frac{T_i-T}{h})y_i}}{\displaystyle\sum_{i=1}^n{K(\frac{T_i-T}{h})}},$$
where $K(\cdot)$ is a kernel function and $h$ is the bandwidth. Let $\tilde{y}_i=y_i-\hat{m}_Y(T_i)$ and $\tilde{X}_i=X_i-\hat{m}_X(T_i)$. Then, in matrix notation, \eqref{formula4} can be rewritten as
\begin{equation}\label{formula5}
\tilde{Y}=\tilde{X}'\beta+{ \bf \epsilon},
\end{equation}
 where $ \tilde{X}=(\tilde{X}'_1 , ...,\tilde{X}'_n)'$ and $\tilde{Y}=(\tilde{y}_1,\tilde{y}_2,...,\tilde{y}_n)'$.
So \eqref{formula5} is a standard linear model, and we may adopt the procedure developed by Zou and Hastie \cite{ZH2005} to study  variable selection for the partially linear model.
 \begin{defn}\label{2-2}
 \ For fixed nonnegative parameters $\lambda_1$ and $\lambda_2$, the Elastic Net procedure for the partially linear model is defined as follows:
 \begin{equation}\label{formula6}
\ L(\lambda_1,\lambda_2,\beta)=\|\tilde{Y}-\tilde{X}'\beta\|^2+\lambda_2\|\beta\|^2+\lambda_1\|\beta\|_1,
\end{equation}
where $\|\beta\|^2=\displaystyle\sum_{j=1}^{p}\beta_j^2$ and $\|\beta\|_1=\displaystyle\sum_{j=1}^{p}|\beta_j|$.

Define
$$\hat{\beta}(\textrm{ENet})=\arg\displaystyle\min_{\beta} L(\lambda_1,\lambda_2,\beta).$$
 \end{defn}

\ According to the Definition 2.1, the Elastic Net procedure becomes Lasso when $\lambda_2=0$ in \eqref{formula6}. By a appropriate transformation, the solution of the Elastic Net procedure can be expressed analogously to the solution form of Lasso (Zou and Hastie \cite{ZH2005}). Thus we can use the least angle regression algorithm (LARS) (Efron et al.\cite{EHJT2004}) to solve it.

One of the key issues is  the choice of the parameters $\lambda_n (n=1,2)$ and $h$. Here we fix  $\lambda_2$ and choose the optimal values of $\lambda_1$ by Cross-validation (Verweij \cite{V1993}). For the selection of bandwidth, its best value is $O(n^{-1/5})$. So we find effective bandwidth $h$ for $\hat{m}_X(T)$ and  $\hat{m}_Y(T)$ with interpolation technique proposed by Ruppert et al.\cite{RSW1995}.

\section{Group effect}
Collinearity is a major obstacle in dealing with high-dimensional data. Eliminating collinearity in the determination of the best linear model is a vital subject. In this section, we  investigate the group effect of the Elastic Net procedure.

\begin{thm}
 Assume that the response $Y$ is centred and the predictor $X$ is standardized. Given the data $({\bf X,Y})$ and parameters $(\lambda_1,\lambda_2)$, let $\hat{\beta}(\lambda_1,\lambda_2)$ be the Elastic Net estimation. Assume that  $\hat{\beta}_k(\lambda_1,\lambda_2)\hat{\beta}_l(\lambda_1,\lambda_2)>0$.
\noindent Define the group effect $D_{\lambda_1,\lambda_2}(k,l)$ by
\begin{equation}\label{formula7}
{D_{\lambda_1,\lambda_2}(k,l)}=\big|\hat{\beta}_k(\lambda_1,\lambda_2)-\hat{\beta}_l(\lambda_1,\lambda_2)\big|.
\end{equation}
Then
\begin{equation}\label{formula8}
{D_{\lambda_1,\lambda_2}(k,l)}\leq\frac{2m}{\lambda_2}\displaystyle\sum_{i=1}^n{|\hat{r}_i|},
\end{equation}
where $m=\displaystyle\max_{i\in\{1,\cdots,n\}}\{{|x_{ik}-x_{il}|}\}$ and $\hat{r}_i$ given by \eqref{a-1} is the predicted residual.
\end{thm}

\noindent{\it Proof:} Since $\hat{\beta}_k(\lambda_1,\lambda_2)\hat{\beta}_l(\lambda_1,\lambda_2)>0$,
$$\textrm{sgn}\{\hat{\beta}_k(\lambda_1,\lambda_2)\}=\textrm{sgn}\{\hat{\beta}_l(\lambda_1,\lambda_2)\},$$ where $\textrm{sgn}\{\cdot\}$ is the sign function.

 Let $\hat{\beta}_m(\lambda_1,\lambda_2)\neq0$.
 Note that $\hat{\beta}(\lambda_1,\lambda_2)$
satisfies $$\frac{\partial L(\lambda_1,\lambda_2,\beta)}{\partial \beta_m}|_{\beta=\hat{\beta}(\lambda_1,\lambda_2)}=0.$$
\noindent Moreover, we have
\beqnn
L(\lambda_1,\lambda_2,\beta)&&=\|\tilde{Y}-\tilde{X}'\beta\|^2+\lambda_2\|\beta\|^2+\lambda_1\|\beta\|_1
\\&&=\sum_i(\tilde{y_i}-\displaystyle\sum_j{\beta_j{\tilde{x}_{ij}}})^2+
\lambda_2\sum_j{\beta_j^2}+\lambda_1\displaystyle\sum_j{|\beta_j|}
\\&&=\sum_i(\tilde{y}_i^2-2\tilde{y_i}\displaystyle\sum_j{\beta_j{\tilde{x}_{ij}}}+(
\displaystyle\sum_j{\beta_j{\tilde{x}_{ij}}})^2)+\lambda_2\displaystyle\sum_j{\beta_j^2}+\lambda_1\sum_j{|\beta_j|}.
\eeqnn
\noindent Therefore,
\beqlb\label{1-1}
\-2\sum_i\tilde{y_i}\tilde{x}_{ik}+2\displaystyle\sum_i\tilde{x}_{ik}
{\sum_j{\hat{\beta}_j(\lambda_1,\lambda_2)}\tilde{x}_{ij}}+\lambda_1\textrm{sgn}\{\hat{\beta}_k(\lambda_1,\lambda_2)\}
+2\lambda_2\hat{\beta}_k(\lambda_1,\lambda_2)=0,
\eeqlb
and
\beqlb\label{1-2}
\ -2\sum_i\tilde{y_i}\tilde{x}_{il}+2\sum_i\tilde{x}_{il}
{\sum_j{\hat{\beta}_j(\lambda_1,\lambda_2)}\tilde{x}_{ij}}+\lambda_1\textrm{sgn}\{\hat{\beta}_l(\lambda_1,\lambda_2)\}
+2\lambda_2\hat{\beta}_l(\lambda_1,\lambda_2)=0.
\eeqlb
By (\ref{1-1}) and (\ref{1-2}), we have
$${\hat{\beta}_k(\lambda_1,\lambda_2)-\hat{\beta}_l(\lambda_1,\lambda_2)}=
 \frac{1}{\lambda_2}\sum_i(\tilde{x}_{ik}-{\tilde{x}_{il}})\big(\tilde{y}_i-\sum_j{\hat{\beta}_j(\lambda_1,\lambda_2)\tilde{x}_{ij}}\big).$$
On the other hand, we have $$\tilde{x}_{ik}=x_{ik}-\hat{m}_X(T_i)_k=x_{ik}-\frac{\displaystyle\sum_j{K(\frac{T_j-T_i}{h})x_{jk}}}{\displaystyle\sum_j{K(\frac{T_j-T_i}{h})}},$$ and
$$\tilde{x}_{il}=x_{il}-\hat{m}_X(T_i)_l=x_{il}-\frac{\displaystyle\sum_j{K(\frac{T_j-T_i}{h})x_{jl}}}{\displaystyle\sum_j{K(\frac{T_j-T_i}{h})}}.$$

Therefore,

\beqnn
&&\Big|{\hat{\beta}_k(\lambda_1,\lambda_2)-\hat{\beta}_l(\lambda_1,\lambda_2)}\Big|
\\&&\qquad\qquad=
\frac{1}{\lambda_2}\sum_i|x_{ik}-x_{il}-{(\frac{\displaystyle\sum_j{K(\frac{T_j-T_i}{h})x_{jk}}}
{\displaystyle\sum_j{K(\frac{T_j-T_i}{h})}}}-{\frac{\displaystyle\sum_j{K(\frac{T_j-T_i}{h})x_{jl}}}
{\displaystyle\sum_j{K(\frac{T_j-T_i}{h})}}})||\hat{r}_i|
\\&&\qquad\qquad\leq\frac{1}{\lambda_2}\sum_i(|x_{ik}-x_{il}|+\frac{\sum_j{K(\frac{T_j-T_i}{h})|x_{jk}-x_{jl}|}}
{\displaystyle\sum_j{K(\frac{T_j-T_i}{h})}})|\hat{r}_i|
\\&&\qquad\qquad\leq\frac{1}{\lambda_2}\displaystyle\sum_i(|x_{ik}-x_{il}|+|x_{jk}-x_{jl}|)|\hat{r}_i|,
\eeqnn
where \beqlb\label{a-1}\hat{r}_i=\tilde{y}_i-\displaystyle\sum_j{\hat{\beta}_j(\lambda_1,\lambda_2)\tilde{x}_{ij}}.\eeqlb

 Let $m=\displaystyle\max_{i\in\{1,2,...,n\}}\{{|x_{ik}-x_{il}|}\}$. Then we have
\ $$\frac{1}{\lambda_2}\displaystyle\sum_i(|x_{ik}-x_{il}|+|x_{jk}-x_{jl}|)
|\hat{r}_i|\leq\frac{2m}{\lambda_2}\displaystyle\sum_i|\hat{r}_i|,$$
that is $$\Big|{\hat{\beta}_k(\lambda_1,\lambda_2)-\hat{\beta}_l(\lambda_1,\lambda_2)}\Big|\leq\frac{2m}{\lambda_2}\displaystyle\sum_i|\hat{r}_i|.$$\qed

$D_{\lambda_1,\lambda_2}(k,l)$ describes the difference between the coefficient paths of predictors $k$ and $l$. $m\rightarrow0$ means $x_k$ and $x_l$ are highly correlated. Then the theorem 3.1 suggests that the difference between the coefficient paths of predictor $k$ and predictor $l$ is almost zero.  If $\hat{\beta}_k(\lambda_1,\lambda_2)\hat{\beta}_l(\lambda_1,\lambda_2)<0$, we consider the $-x_k$. The upper bound in  \eqref{formula8} provides quantitative description for the group effect of the Elastic Net. It can be seen that the Elastic Net procedure has the ability to do group selection, but the Lasso fails (Efron et al.\cite{EHJT2004}).

\section{Simulation study}
In this section we report a numerical simulation study to compare the Elastic Net procedure with  Lasso, ALasso, and Ridge. We have known that all the four methods can deal with collinearity problems well. However, the last three methods can only select one of the highly correlated predictors. As the statement in the theorem, all the necessary highly correlated variables can be selected into the model by the Elastic Net procedure. In the extreme situation where some variables are exactly identical, the last three methods can only select one of the identical variables into the model. But all the identical variables can be selected into the model  by the Elastic Net procedure. Moreover, it can assign identical coefficients to the identical variables. We now demonstrate the above argument by the following numerical simulation.

We generated data from the partially linear model: $Y=X'\beta+f(T)+\varepsilon$, where $\beta=( -2,1,1,0,2/3,0,0,0)'$, $X=(x_1,x_2,x_3,x_4,x_5,x_6,x_7,x_8)'$, $f(T)=T^2$ with $T\sim{U[-1,1]}$, and $\varepsilon\sim{N(0,0.04)}$. Moreover, we assume that $x_3=x_2$, $x_4=\frac{2}{3}x_1+\frac{1}{3}x_2+\frac{1}{3}x_3+\frac{2}{3}e$, where $x_i,i=1,2,5,6,7,8$ and $e$ follow $N(0,1)$. The kernel function is \beqnn K(y)=\begin{cases}\frac{1}{2},&|y|\leq1, \\ 0,& \textrm{others}.\end{cases}
 \eeqnn We did the simulations   for $n=1000$ and repeated 50 times by using the software $R$. We considered the Lasso, ALasso, Ridge and the Elastic Net procedure for the variable selection. We turned ALasso and Elastic Net procedure into Lasso and estimate coefficients by LARS. We picked a value for $\lambda_2$, say $\lambda_2=1/3$. We chose the optimal values of the parameters $\lambda_1$ by 10-fold CV. The best value of bandwidth is $O(n^{-1/5})$. So we found effective bandwidth $h$ for $\hat{m}_X(T)$ and  $\hat{m}_Y(T)$ with interpolation technique. The coefficients estimates are in Table 1. The MSE (mean squared error) are in Table 2, where $MSE=\|\hat{\beta}-\beta\|^2.$
\begin{table}
\centering
\caption{The mean value of coefficient estimates based on 50 replications.}
\begin{tabular}{l|cccccccc}
\hline
\ Var &$x_1$&$ x_2$ & $x_3$ &$x_4$ &$x_5$ & $x_6$ & $x_7$ & $x_8$\\
\hline
Lasso & -1.81126 &1.69923&0&0 &0.59680&-0.00024&0.00697&-0.00095\\
ALasso&-1.86473&1.76253&0&0&0.63182&0&0.00002&0\\
Ridge&-1.33509&1.67174&0&0.00151&0.70761&0.07696&0.00711&0.14399\\
ENet&-1.85062&0.73087&0.73087 &0&0.57687&-0.00032&0.00036&-0.00011\\
\hline
\end{tabular}
\end{table}
\begin{table}
\centering
\caption{The MSE of the methods based on 50 replications.}
\begin{tabular}{l|cccccccc}
\hline
Methods&Lasso&ALasso&Ridge&ENet\\
\hline
MSE& 1.529
&1.601&1.921&0.175\\
\hline
\end{tabular}
\end{table}

Several observations can be made from  Tables 1 and 2. The last three methods can only select the variable $x_2$. Both $x_2$ and $x_3$ are selected to the model by the Elastic Net procedure. The Elastic Net procedure can assign identical coefficients to the identical variables. By using ALasso, we got that $x_6$ and $x_8$ are out of the model and $x_7$ is almost zero. The zero components can be eliminated more correctly to the final model by ALasso for its oracle property than other methods. The Ridge almost selects all the variables into the model. The Elastic Net procedure can select all the highly correlated variables into the model accurately. We can see from the results that the Elastic Net procedure works better than the other three methods  in dealing with the data of strongly correlated variables.

\section{Real data example}
 A typical microarray data set has thousands of genes and less than 100 samples. Because of the unique structure of the microarray data, a good variable selection method should have the following properties:
\begin{enumerate}
\item[(1)] Gene selection should be built into the procedure.
\item [(2)]It should not be limited by the fact that ${p} \gg {n}$.
\item [(3)] For those genes sharing the same biological pathway, it should be able to automatically include whole groups into the model once one gene among them is selected.
\end{enumerate}

Most of  popular methods fail with respect to at least one of the above properties (Zou and Hastie\cite{ZH2005}). The Lasso is good at (1) but fails to (2) and (3). As an automatic variable selection method, the Elastic Net procedure naturally overcomes the difficulty of ${p}\gg {n}$ and has the ability to do group selection. We use the leukemia data to illustrate the advantage of the Elastic Net procedure for partially linear models.

The leukemia data consists of 3571 genes and 72 samples (Golub et al. \cite{Getal1999}). In the training data set, there are 38 samples, among which 27 are
type 1 leukemia (ALL) and 11 are type 2 leukemia (AML). The goal is to construct a diagnostic rule based on the expression level of those 3571 genes
to predict the type of leukemia. The remaining 34 samples are used to test the prediction accuracy of the diagnostic rule. To apply the Elastic Net, ALasso and Lasso, we first coded the type of leukemia as a 0-1 response $y$. We did the variable selection by Lasso, ALasso and Elastic Net. The kernel function $K(\cdot)$ is the same as in the Sec.4. We used 10-fold CV to select the tuning parameters.
\begin{table}
\centering
\caption{Summary of the leukaemia data.}
\begin{tabular}{lcccc}
\hline
Method&Test error&No. of genes& Step of selection\\
\hline
Lasso&4/34&26&29\\
\hline
 ALasso&3/34&22&25\\
\hline
ENet&2/34&51&60\\
\hline
\end{tabular}
\label{Table3}
\end{table}

\begin{figure}
\centering
\subfigure[The coefficients paths at each step of Lasso.] {\includegraphics[height=7cm,width=7cm]{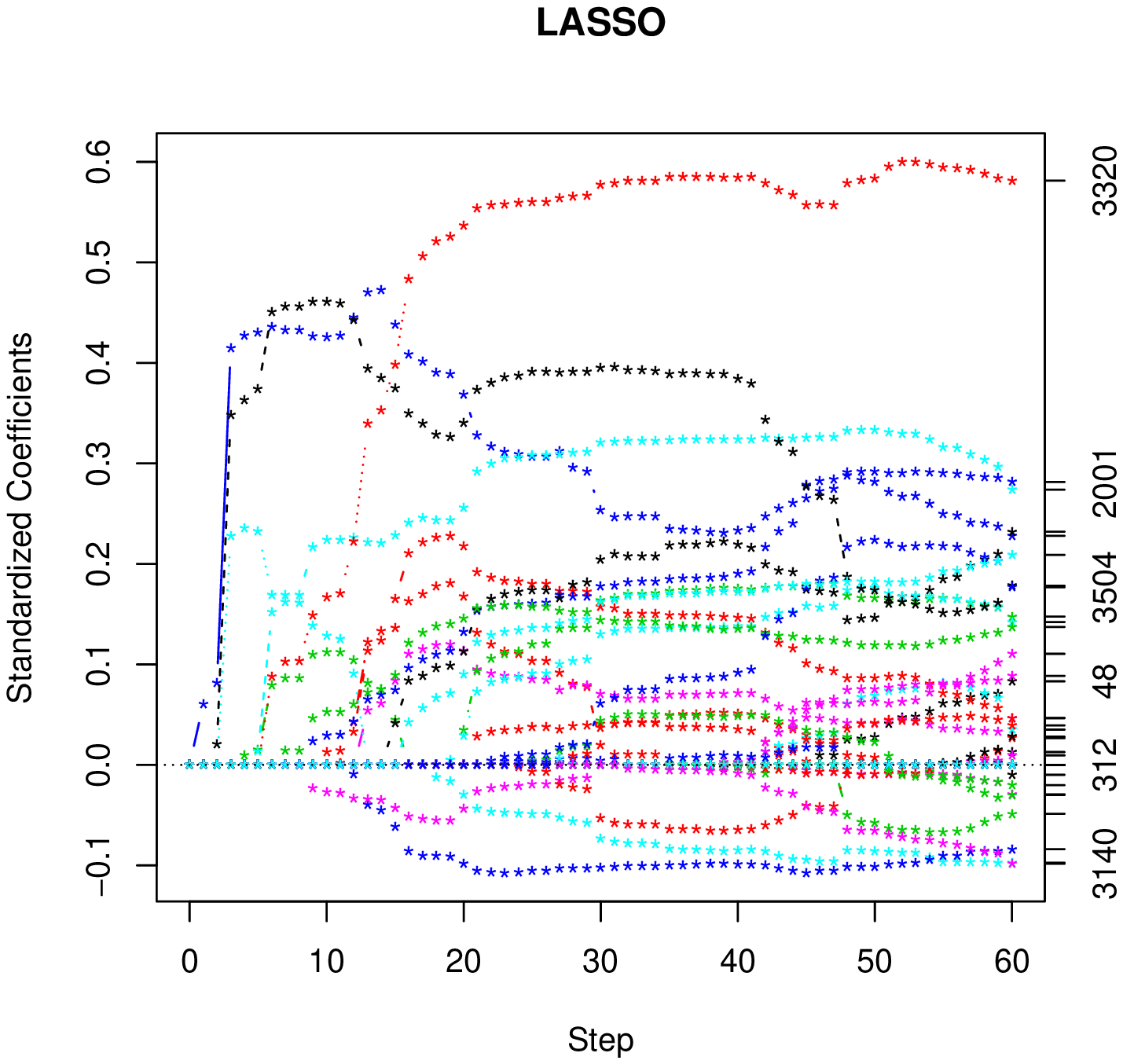}}
\subfigure[The coefficients paths at each step of ALasso.] {\includegraphics[height=7cm,width=7cm]{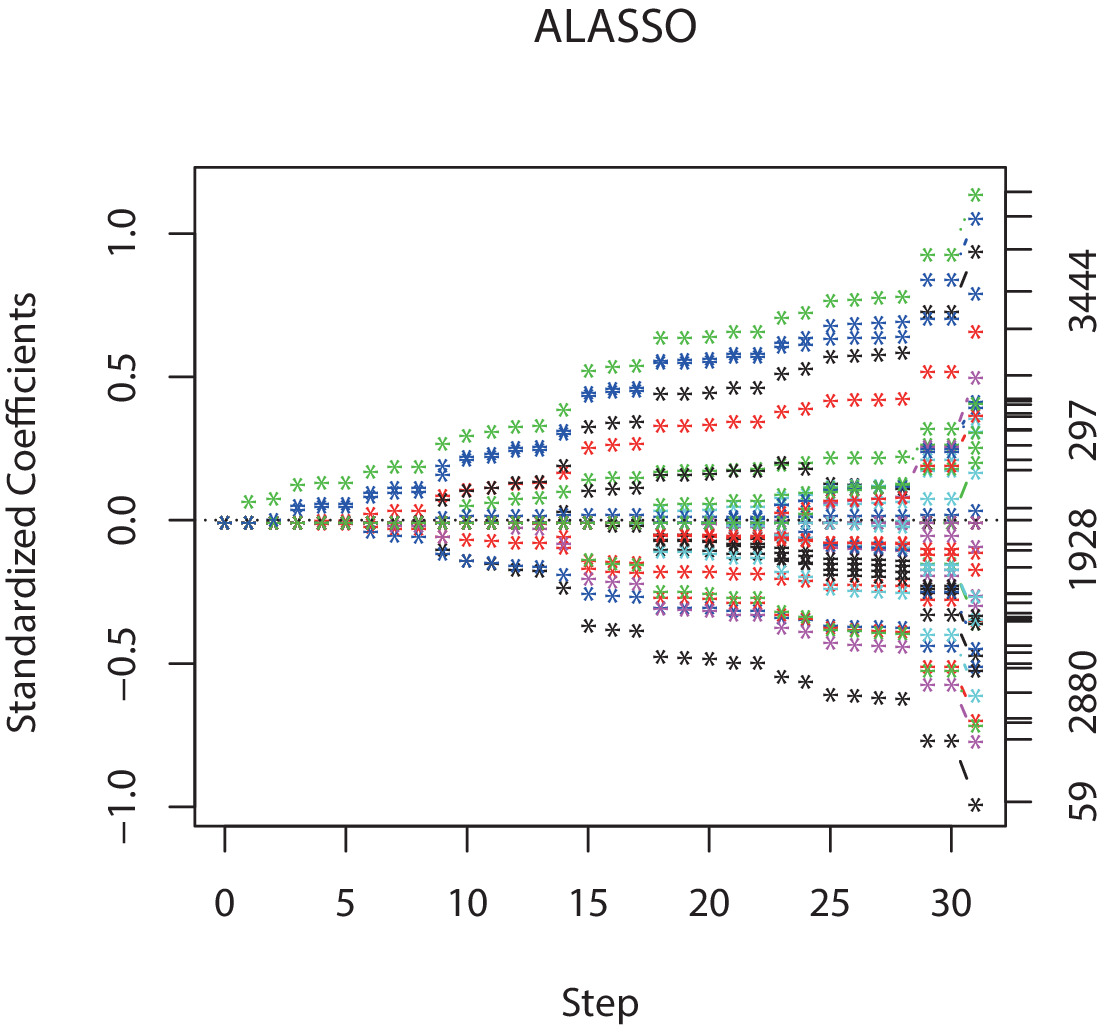}}
\subfigure[The coefficients paths at each step of ENet.] {\includegraphics[height=7cm,width=7cm]{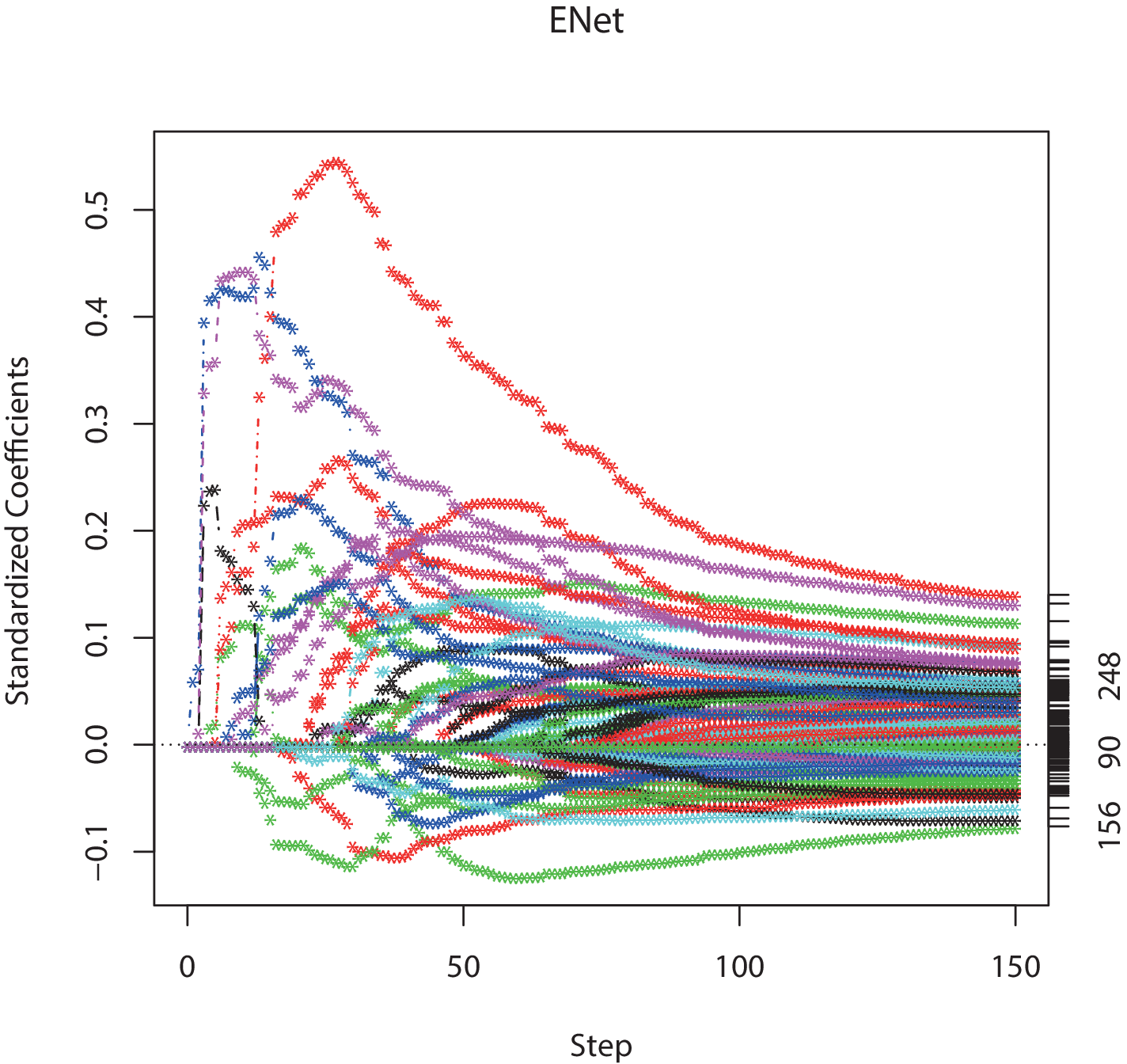}}
\caption{The coefficients paths at each step of Lasso, ENet and ALasso.}
\label{fig1}
\end{figure}
We stopped the Lasso after 60 steps,  ALasso after 30 steps, and the Elastic Net after 150 steps. Table \ref{Table3} compares the Elastic Net with Lasso and ALasso. The Elastic Net gives the better classification, and it has an internal gene selection facility. Figure \ref{fig1} displays the solution paths and the gene selection results. We get that the number of genes selected into the model by Lasso is 26 at step 29, while the ALasso is 22 at step 25. The zero components can be eliminated to the final model by ALasso for its Oracle property than Lasso.  The optimal Elastic Net model is given at step 60 with 51 selected genes. Note that the size of the sample is 38, so the Lasso and ALasso can at most select 38 genes. In contrast, the Elastic Net selects more than 38 genes, not limited by the sample size. The Elastic Net is particularly useful when the number $p$ of predictors  is much bigger than the sample size $n$. Neither Lasso nor ALasso is  a very satisfactory variable selection method in the  case $p \gg n$.

\section{Conclusions}

Collinearity between variables is a problem we usually encounter in high-dimensional data. If it can not be handled properly, the accuracy of models we get does not reach the standard required and it will affect the interpretability of the models seriously.  In this paper, we have proposed a more effective selection method,  Elastic Net procedure,  to eliminate the collinearity and select all the strongly correlated variables. The Elastic Net procedure for partially linear models produces a sparse model with good prediction accuracy, while encourages a group effect. The simulations and empirical results demonstrate the good performance of the Elastic Net and its superiority over the other methods.

\medskip
\noindent{\bf Acknowledgments:}\ The authors thank two anonymous referees for very detailed comments and suggestions. This work was  supported by the Natural Science Foundation of China (No.11361007), the Guangxi Natural Science Foundation (Nos.2012GXNSFBA053010 and 2014GXNSFCA118001) and  the Project for Fostering Distinguished Youth Scholars of Shandong University of Finance and Economics.

\end{document}